\documentclass[%
reprint
superscriptaddress,
amsmath,amssymb,
aps,
prb,
twocolumn,
]{revtex4-2}

\usepackage{graphicx}
\usepackage{dcolumn}
\usepackage{bm}
\usepackage{multirow}
\usepackage{float}
\usepackage{overpic}
\usepackage{subcaption}
\usepackage{caption}
\usepackage{makecell}
\usepackage{oplotsymbl}
\usepackage{todonotes}
\usepackage{hyperref}
\usepackage{color}
\usepackage{graphpap}
\usepackage{ulem}

\begin{document}



\title{Location and migration of interstitial Li ions in CsPbI$_3$ crystals}


\author{Wei Wei}
\affiliation{Freiburg Center for Interactive Materials and Bioinspired Technologies, University of  Freiburg, Georges-K\"ohler-Allee 105, 79110 Freiburg, Germany}
\affiliation{Fraunhofer Institute for Mechanics of Materials IWM, W\"ohlerstra{\ss}e 11, 79108 Freiburg, Germany}

\author{Julian Gebhardt}
\affiliation{Fraunhofer Institute for Mechanics of Materials IWM, W\"ohlerstraße 11, 79108 Freiburg, Germany}
\affiliation{Freiburg Center for Interactive Materials and Bioinspired Technologies, University of Freiburg, Georges-K\"ohler-Allee 105, 79110 Freiburg, Germany}

\author{Daniel F. Urban}
\affiliation{Fraunhofer Institute for Mechanics of Materials IWM, W\"ohlerstra{\ss}e 11, 79108 Freiburg, Germany}
\affiliation{Freiburg Materials Research Center, University of Freiburg, Stefan-Meier-Stra{\ss}e 21, 79104 Freiburg, Germany}

\author{Christian Elsässer}
\email{christian.elsaesser@iwm.fraunhofer.de}
\affiliation{Fraunhofer Institute for Mechanics of Materials IWM, W\"ohlerstra{\ss}e 11, 79108 Freiburg, Germany}
\affiliation{Freiburg Materials Research Center, University of Freiburg, Stefan-Meier-Stra{\ss}e 21, 79104 Freiburg, Germany}
\affiliation{Freiburg Center for Interactive Materials and Bioinspired Technologies, University of Freiburg, Georges-K\"ohler-Allee 105, 79110 Freiburg, Germany}

\date{\today}

\begin{abstract}

Halide perovskites are highly promising light-harvesting materials with strong ionic \sout{bonding} character, enabling in principle the combination of a solar cell and a Li-ion battery in one integrated photo-battery device. Here, we investigate Li ions inside crystals of CsPbI$_3$, as a prototype compound, by means of density-functional-theory calculations. Our findings demonstrate that the interstitial location and migration of Li ions depend strongly on the dynamic nature of the crystal structure of the perovskite compound. We consider two limiting cases for Li in CsPbI$_{3}$,(i) the cubic-symmetry structure as a model for the limit of fast ion motion and (ii) a distorted cubic structure as a model for the limit of slow ion motion. For both limiting cases we obtain moderate energy barriers for migrating Li ions, which highlight the potential of halide perovskites like CsPbI$_3$ for applications in photo-battery devices.

\end{abstract}


\maketitle

\section{Introduction}

The concept of photo-charging batteries using halide perovskites as the light-absorbing material is attracting a lot of attention in recent years \cite{zhang2020halide}. The halide perovskites have emerged as very promising absorber materials in the field of photovoltaic energy conversion 
\cite{Jena2019,Zhao2019,Yin2015,Tian2020,Xiang2019,Cui2015}, offering both high photovoltaic conversion efficiency and low fabrication cost \cite{NREL,he2021wide}. Furthermore, unlike most traditional photovoltaic materials, organic-inorganic hybrid halide perovskites are generally considered to be mixed ionic-covalent character, and it is this mixing of properties that allows both electronic and ionic conduction \cite{walsh2015principles,du2014efficient，kuku1987electrical,kuku1998ionic}. Initial research concerning the ionic character of these materials primarily focused on the side effects associated with internal ionic migration, i.e., hysteresis effects and unstable power conversion efficiency as well as degradation processes associated with these materials \cite{guerrero2016interfacial,domanski2016not,zhang2020halide}. However, halide perovskites were recently reported to have good ionic conductivity and storage potential of interstitial Li ions, thereby enabling, in principle, the integration of a solar cell and a Li-ion battery into a single photo-battery device \cite{li2017extrinsic}.

The thermodynamical stability of halide perovskites and the concentration limit of interstitial Li ions are intensively discussed \cite{vicente2017methylammonium,dawson2017mechanisms,buttner2022are}. The literature provides recent studies covering a wide range of Li intercalation concentrations, and the intercalation mechanism is not yet fully explored \cite{dawson2017mechanisms,buttner2022are,zhang2020halide}. According to the literature, organic-inorganic hybrid halide perovskites are less suitable than all-inorganic halide perovskites for use in Li intercalation type photobatteries and their experiments have shown decomposition \cite{dawson2017mechanisms,buttner2022are}. Dawson et al. show that conversion reactions are energetically preferable to Li intercalation in MAPbI$_3$ \cite{dawson2017mechanisms}. To solve the problem, some efforts focus on all-inorganic perovskites. One of the promising compounds is CsPbI$_3$ due to its moderate band gap, and the compound is chemically more stable against decomposition \cite{zhou2019chemical}. Though the most stable phase of this compound is a non-perovskite-type $\delta$ phase \cite{marronnier2018anharmonicity,straus2020understanding}, Eperon et al. found an experimental method to maintain CsPbI$_3$ stability in its black phase at room temperature and realized the first working CsPbI$_3$ solar cell \cite{eperon2015inorganic}. The Li intercalation in CsPbI$_3$ perovskite is less studied, but the coexistence of Li ions in other all-inorganic halide perovskites are reported in the literature \cite{zhang2020halide}. For instance, Jiang et al. show Li doping in CsPbBr$_3$ with a concomitant photoluminescence blue shift \cite{jiang2017electrochemical}. Thus, we propose that the Li intercalation in solid state materials system with all-inorganic halide perovskites is feasible, at least in such a low Li concentration as we have in our study. 

Li ions can be put into some halide perovskites, but open questions remain concerning the mechanism of Li-ion migration through the perovskite structure of a potential photo-battery material. Ionic migration was detected in halide perovskite materials by various electrical measurements performed on operating solar cells, such as chronoamperometry, capacitance, or impedance methods \cite{lopez2018device,yang2015significance,peng2018quantification}. But explorations of the migration behavior of Li ions are limited. Some studies have reported diffusivity values for natural defects in halide perovskites similar to those encountered in conventional solid-state ion conductors \cite{eames2015ionic,richardson2016can}, while other analyses have indicated even faster ion migration \cite{yang2015significance,almora2016ionic}. Despite the recognition of the need to understand the kinetics of ion migration, the complex defect chemistry of halide perovskites makes it challenging to discern the migration rates of specific mobile ions within the perovskite materials itself \cite{vicente2017organohalide}.

The migration of interstitial Li ions is highly dependent on the structure of the host crystal. Concerning specifically the halide perovskites, some initial studies have examined the cubic perovskite phase ABX$_3$ due to its excellent photovoltaic properties, and treated its crystal structure as a high-symmetry static crystal \cite{afsari2016electronic,jong2018first}. However, instead of being a static crystal, the current understanding of the cubic phase of halide perovskites is that of a spatial and temporal average of a dynamic finite-temperature structure \cite{wiktor2017predictive,gebhardt2021efficient}. 
Hence, Li-ion diffusion models need to cope with dynamic structures of halide perovskites.

In this work, we use CsPbI$_3$, a well-studied inorganic halide perovskite, as a prototype to understand the location and migration of interstitial Li ions in a set of static approximants for the dynamic perovskite structure. By means of density-functional-theory (DFT) calculations, we investigate two scenarios: (i) Li migration in the high-symmetry cubic perovskite structure, and (ii) Li migration in a perovskite structure with tilted corner-shared Pb-I octahedra. Taking into account possible transition states between crystal structures, these two model systems serve as limiting cases for estimating the dynamical behavior of migrating Li ions at finite temperatures. Our results provide an adequate understanding of the Li-ion migration in the CsPbI$_3$ single crystal, and we believe that the results can be qualitatively extrapolated to some extent to other halide perovskites, at least to inorganic ones.

The paper is organized as follows. In Sec.\ \ref{sec:theory} we describe the theoretical framework underlying our atomistic simulations by introducing the studied structure models and the computational setup. In Sec.\ \ref{results} we present our results obtained for the models of the fast ion limit and the slow ion limit. Then we discuss the stability and mobility of interstitial Li ions in Sec.\ \ref{discussion} and compare our results with other Li-ion battery materials. Finally we summarize our major findings and conclusions in Sec.\ \ref{summary}.

\section{Theoretical Approach}
\label{sec:theory}

\subsection{Crystal phases of CsPbI$_3$}

CsPbI$_3$ is known to exist in four crystalline
phases, namely, the $\alpha$ ($Pm\overline{3}m$), $\beta$ ($P4/mbm$), $\gamma$ ($Pnma$), and $\delta$ ($Pnma$) phases \cite{eperon2015inorganic}.
The first three are referred to as the perovskite phases. 
The crystal structure of the $\delta$-phase
is very different from the perovskite structures.
The $\delta$ phase is thermodynamically favored at ambient
conditions \cite{eperon2015inorganic}, but with modern experimental thin-film
or nano-crystal technologies it is possible to synthesize CsPbI$_3$
in the $\alpha$, $\beta$, and $\gamma$ perovskite phases at ambient conditions, too \cite{marronnier2017structural, eperon2015inorganic,mahato2020highly}.
These three phases have recorded electronic band gaps of 1.73~eV \cite{eperon2015inorganic}, 1.94~eV and 2.03 eV \cite{mahato2020highly}, which are all in the proper range for absorption of visible light.
Experimentally, at elevated temperatures the cubic $\alpha$ phase is the most stable structure
among the three perovskite phases. Upon cooling, phase transitions are observed at 533 K and
448 K towards the tetragonal $\beta$ phase and the orthorhombic $\gamma$ phase \cite{marronnier2017structural}, respectively. 

The cubic crystal structure of the $\alpha$ phase of CsPbI$_3$ is observed in diffraction experiments as a temporal
and spatial average \cite{whitfield2016structures,yang2017spontaneous} of atomic positions. It has been used as a static structure model in most
previous theoretical studies \cite{zhang2019improved,kye2019vacancy,jong2018first}. As outlined in Refs.\ \cite{gebhardt2021efficient,whitfield2016structures,wiktor2017predictive,yang2017spontaneous}, it is
unlikely that the static $\alpha$-phase structure is an appropriate representation of the dynamical
CsPbI$_3$ crystal around room temperature, and it is essential to consider the dynamic vibrations of the perovskite structure. The conventional $\alpha$ and $\beta$ phases 
are now understood to result from dynamical averaging of lower symmetry configurations at finite temperatures \cite{marronnier2017structural,wiktor2017predictive,gebhardt2021efficient}. We will demonstrate below that this dynamics becomes even more relevant when adding Li to the structure.

In our discussion of Li migration in CsPbI$_3$ we will frequently make use of the pseudo-cubic variants of the $\beta$ and $\gamma$ phases, i.e. structures for which the lattice parameters $a$, $b$, and $c$ are kept equal while allowing a tilting of the Pb-I octahedra.
Starting from the cubic $\alpha$ phase, we keep the cubic shape of the $2\times 2\times 2$ supercell fixed and create a structure
analogous to the $\gamma$ structure by tilting the octahedra by angles $\theta=13.9^\circ$ and
$\eta=12.7^\circ$, cf.\ Fig.\ \ref{fig_all_phase}. For this tilted structure, the volume is optimized by minimizing its energy
while fixing the three edges of the supercell at the same length. The resulting structure is
denoted as $\gamma'$ and has $Pnma$ symmetry. In this pseudo-cubic cell, we generate structures
$\beta'$ and $\alpha'$ as analogues to $\beta$ and $\alpha$, by reverting the respective introduced tilts of octahedra. 
Moreover, we introduce the $\beta_\eta'$ structure with only a non-zero tilt angle $\eta=13.1^\circ$.
Table \ref{tab:phases} compiles the structural parameters
and relative energies of all the structures considered in this work. We observe that the $\gamma'$ structure, which has both non-zero tilt angles $\theta$ and $\eta$, has the lowest energy among the four pseudo-cubic structures. The $\alpha'$ structure with
the highest symmetry has the highest energy. The $\beta’$ and $\beta_\eta'$ structures, which both have
only one non-zero tilt angle, have similar energies that lie between those of the $\alpha'$ and the $\gamma'$ structures.

\begin{table}[]
\begin{tabular}{cccccccc}
\hline \hline
Phase     &$\alpha$&$\beta$&$\gamma$&$\alpha'$&$\beta'$&$\beta_\eta'$&$\gamma'$\\
\hline
a [\AA]          & 12.79 & 12.52 & 12.36 & 12.64 & 12.64 & 12.64 & 12.64 \\
b/a                & 1.    & 1.    & 1.040 &  1.   &  1.   &  1.   & 1. \\ 
c/a                & 1.    & 1.036 & 1.025 &  1.   &  1.   &  1.   & 1. \\
$\theta$ [$^\circ$]& 0.	   & 14.2  & 14.3  &  0.   & 13.0  &  0.   & 13.9 \\
$\eta$ [$^\circ$]  & 0.	   & 0.	   & 12.0  &  0.   & 0.    & 13.1  & 12.7 \\
$\Delta E$ [eV/f.u.] &0.11 & 0.04  & 0.  &  0.13 & 0.06  & 0.04  & 0. \\
\hline \hline
\end{tabular}
\caption{Lattice parameter $a$, lattice-parameter ratios $b/a$ and $c/a$, and bond angles of the different perovskite structures considered in this work. The values refer to a 40-atom $2\times 2\times 2$ perovskite supercell in all cases. The last row lists the difference in total energy per formula unit, as obtained by DFT-PBE calculations.}
\label{tab:phases}
\end{table}

\subsection{Static models for Li in CsPbI$_3$}

For our investigation of the insertion of Li ions into the halide perovskite CsPbI$_3$ by DFT simulations, we consider the $\alpha$ phase of CsPbI$_3$. Its crystal structure has been widely discussed, and two models have been proposed. The
first model considers this phase to be a static high-symmetry cubic perovskite structure \cite{afsari2016electronic,jong2018first}, while
the second model considers it as a dynamic superposition of lower-symmetry structures with tilted Pb-I octahedra \cite{wiktor2017predictive,gebhardt2021efficient}. Note that the $\alpha$ phase retains a pseudo-cubic structure during vibration, so the $\gamma$' structures are better models than the $\gamma$ phase to represent the lower-symmetry structure. 
Following the first model we fixed the host-crystal atoms in the $\alpha$ structure and
allowed Li atoms to migrate between interstital sites in this cubic perovskite crystal. We named this model the fast
ion limit, assuming that the heavier host-crystal atoms cannot follow the fast migration
of the lighter Li atoms and retain the $\alpha$ structure. Following the second
model we allowed Li atoms to fully interact with the surrounding host-crystal atoms in the lower-symmetry $\gamma$' structure.
We named this model the slow ion limit, supposing that all heavy host-crystal atoms have
enough time to relax into equilibrium positions along the paths of the slowly migrating light Li atoms in the $\gamma$' structure.

\begin{figure}
	\includegraphics[width=0.46\textwidth]{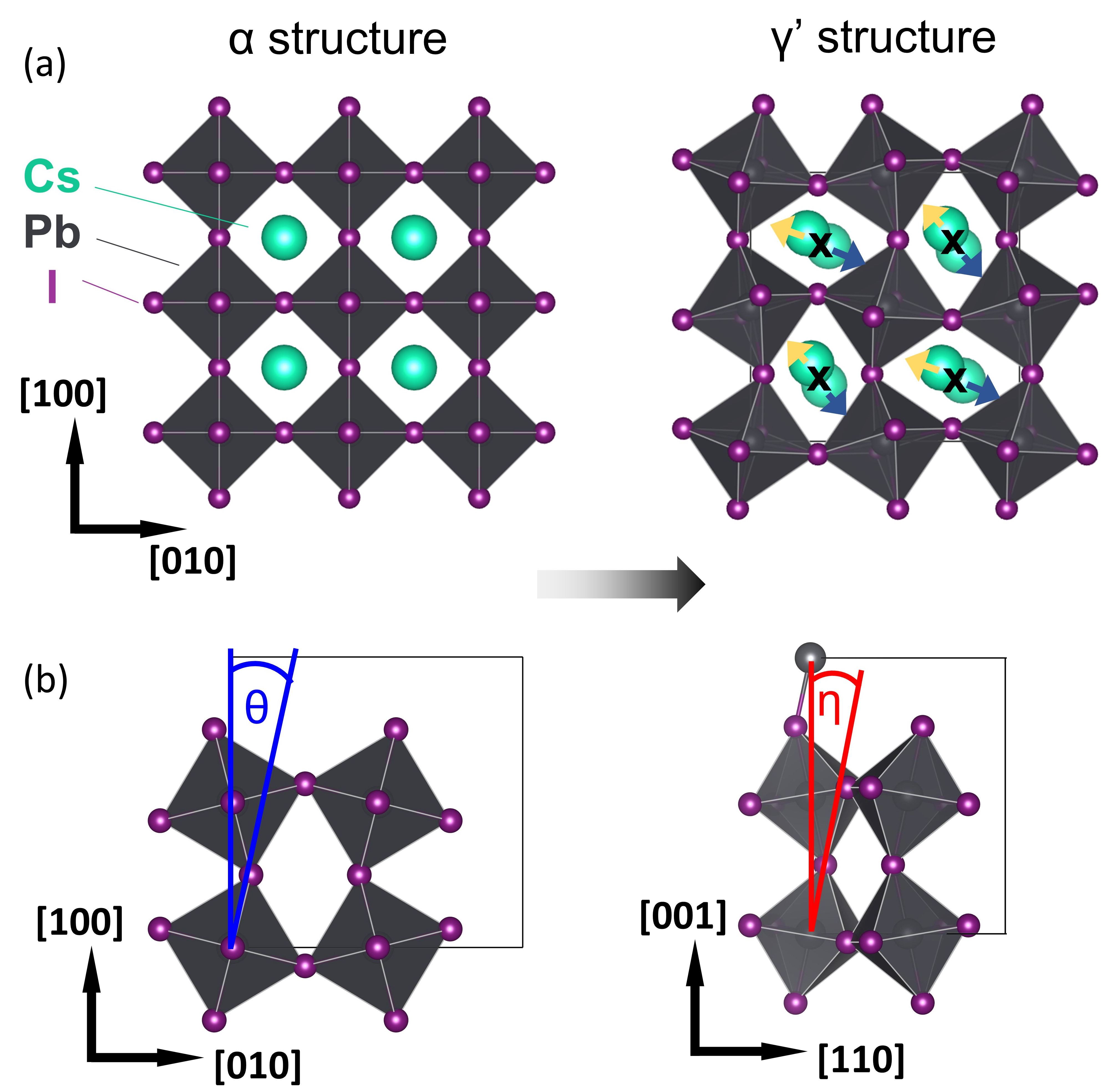}
	\caption{(a) The two limiting models of the dynamic structure of CsPbI$_3$, illustrating the degrees of structural distortion. In the sketch of the $\gamma$' structure the black crosses mark the Cs sites in the $\alpha$ structure, and the arrows indicate the Cs off-center displacements. (b) The tilts of the Pb-I octahedra from $\alpha$ to $\gamma$' are described by the angles $\theta$ and $\eta$.
	}
	\label{fig_all_phase}
\end{figure}

Figure \ref{fig_all_phase} illustrates these two models. The fast ion limit corresponds to the $\alpha$ structure with $Pm\bar{3}m$ symmetry, while the slow ion limit corresponds to the $\gamma$' structure with $Pnma$ symmetry, as introduced in the previous section. 
Both structures are represented by cubic cells, where the Pb atoms are located at the nodes of a (pseudo)cubic lattice. 

\subsection{Computational Details} \label{computational}


All DFT calculations were carried out with the Vienna Ab-initio Simulation Package \cite{kresse1996efficiency} employing projector-augmented-waves pseudopotentials \cite{blochl1994projector}, and the semi-local Perdew-Burke-Ernzerhof (PBE) exchange-correlation functional \cite{perdew1996generalized}. An energy cutoff of 520 eV was used for the plane-waves basis. Total-energy differences and forces on atoms for all structural degrees of freedom are converged within 1$\times$10$^{-5}$~eV and 5$\times$10$^{-3}$~eV~\AA$^{-1}$, respectively. 
The Brillouin-zone integrals were sampled by $4\times 4\times 4$ Monkhorst-Pack $k$-point grids \cite{monkhorst1976special} with a Gaussian smearing of 1$\times$10$^{-3}$~eV for the (2$\times$2$\times$2) supercell models, containing 40 atoms (i.e.\ eight ABX$_3$ formula units). Structural relaxations of the crystals were carried out for all internal coordinates \cite{bucko2005geometry}. The relaxation retains the initially given symmetry of the crystal. The minimum energy paths (MEPs) for the migration of Li ions were calculated by the climbing-image--nudged-elastic-band (CI-NEB) method \cite{henkelman2000climbing} using three images between initial and final state and identical convergence criteria as for the structural relaxation. For the investigation of migrating Li ions in the fast ion limit, the MEP is approximated by computing energies for Li displaced along the straight migration path connecting initial and final state. This simple procedure was compared to a full CI-NEB calculation for the most relevant transition paths, yielding differences in energy barriers that were smaller than 0.01 eV.

Absorption energies of Li ions located at interstitial sites in a perovskite crystal are calculated as 
\begin{equation}
E_{\rm abs}=\frac{E_{\rm tot}[{\rm Li_xCsPbI_3}] - E_{\rm tot}[{\rm CsPbI_3}] - x_{\rm Li} E_{\rm tot}[{\rm Li}_{\rm bcc}]}{x_{\rm Li}}
\end{equation}
where $E_{\rm tot}[{\rm Li_xCsPbI_3}]$ is the total energy of the considered perovskite crystal containing Li, $E_{\rm tot}[{\rm CsPbI_3}] $ is the energy of the reference crystal without Li, $E_{\rm tot}[{\rm Li}_{\rm bcc}]$ is the energy per Li atom in the body-centered cubic one-atom unit cell of the elemental Li metal, and $x \rm _{Li}$ is the proportion of Li per formula unit of the perovskite crystal (we use $x$=1/8 for all cases). The energy of the Li metal was calculated using a k-point grid of 10$\times$10$\times$10 k-points and otherwise unchanged computatinal settings.

\section{Results} \label{results}

\subsection{Structural transitions}

A model of the $\gamma'$ structure can be set up in 24 symmetry-equivalent ways, depending on the crystalographic directions that are used to define the tilting of the Pb-I octahedra. In a given reference coordinate system, the angle $\theta$ can be chosen positive or negative with respect to [001], [010], or [001]. 
The angle $\eta$ can subsequently be defined with respect to any of the four diagonal axes [110], [-110], [1-10], or [-1-10]. 

\begin{figure}
\setlength{\unitlength}{1mm}
\begin{center}
\begin{picture}(85,114)(0,0)
\put(0,0){\includegraphics[width=\columnwidth]{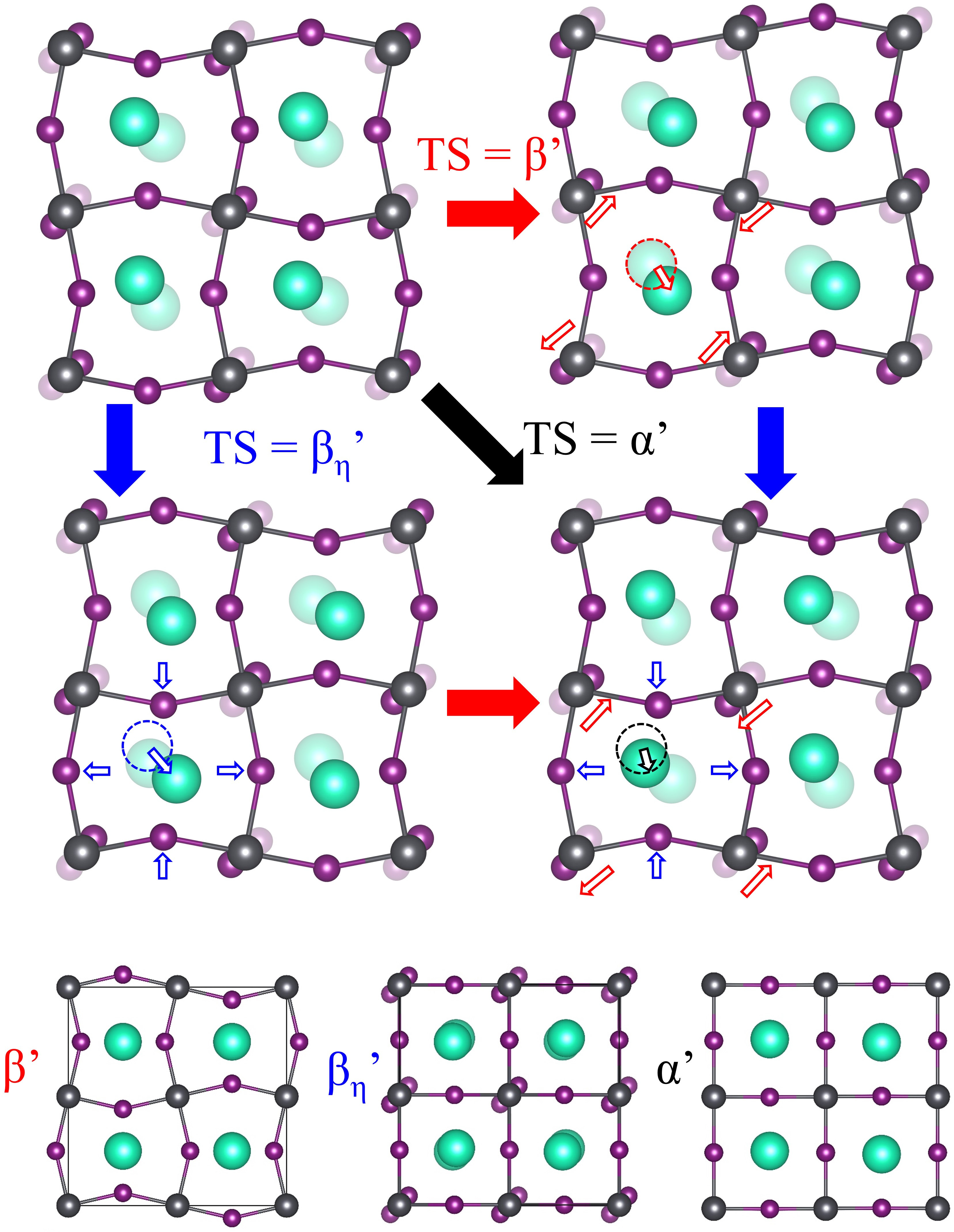}}
\put(0,27){(b)}
\put(0,111){(a)}
\end{picture}
\end{center}
\caption{
	(a) Possible transitions between the 24 symmetry-equivalent configurations of the $\gamma$' structure. 
	In one of the eight formula-unit cells we indicate shifts of atoms accompanying the respective transition by colored arrows:  i) inversion of $\eta$ by moving iodine atoms (red arrows, top right panel), ii) inversion of $\theta$ by moving iodine atoms (blue arrows, lower left panel), or iii) the combination of the former two plus the inversion of the shift of Cs ions (black arrow, lower right panel). The dashed circles indicate the initial positions of the Cs ions. (b) Transition states (TS) that correspond to these three transformations.
}\label{fig_stru_shift}
\end{figure}

\begin{figure}
	\begin{center}
	\includegraphics[width=0.8\columnwidth]{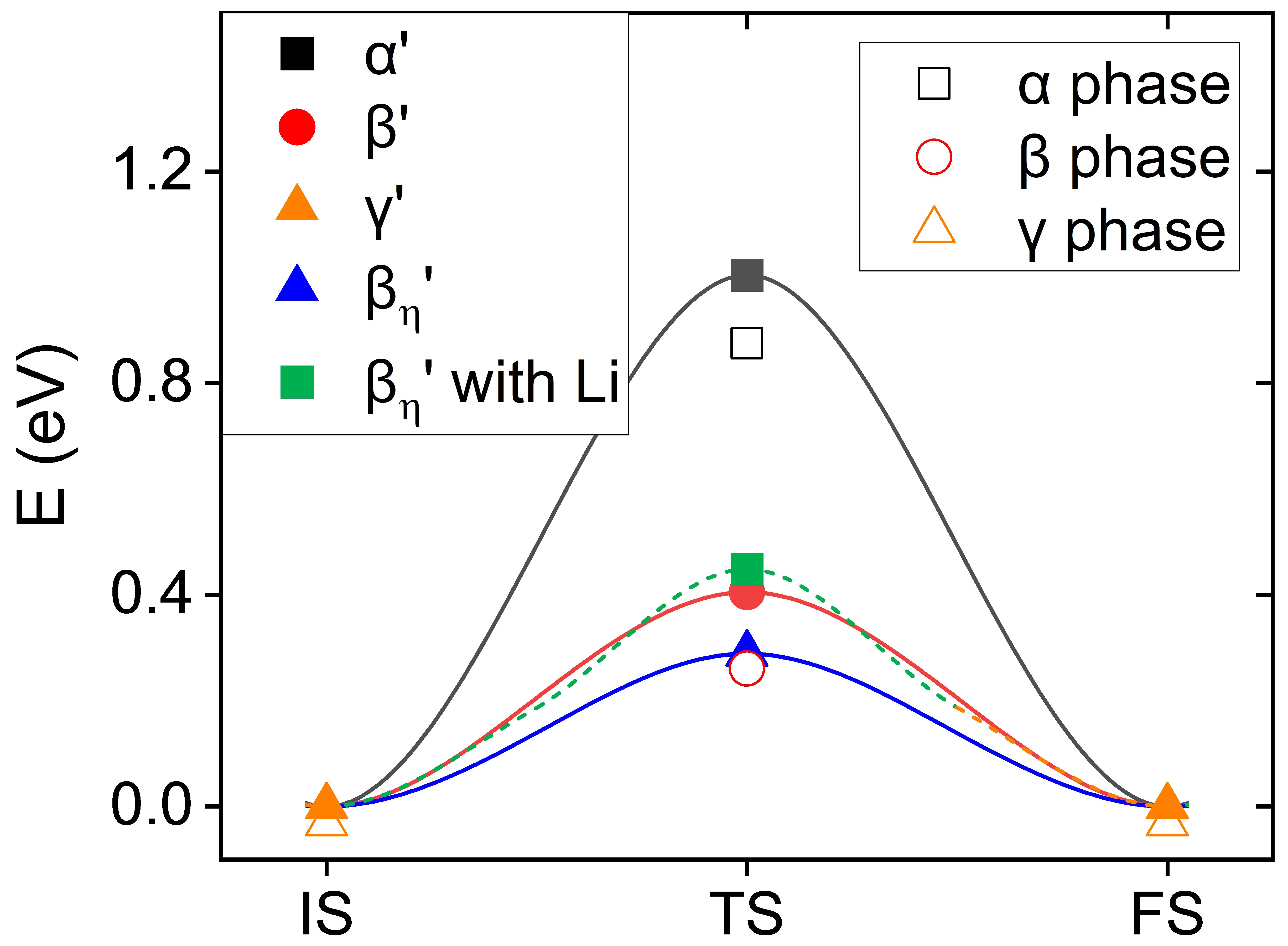}
	\end{center}
	\caption{	
	  Energy barriers for the transitions between two symmetry-equivalent $\gamma$' structures. Empty symbols denote the fully optimized equilibrium structures. The energy differences between these structures and their corresponding structures marked with primes (') are primarily attributed to volume changes and orthorhombic ($\gamma$) or tetragonal ($\beta$) distortions. Energies are referenced with respect to the total energy of the $\gamma$’ structure. (The x-axis labels IS, TS, and FS denote the initial, transition, and final states of the MEPs, respectively.)}
	\label{fig_TS_energies}
\end{figure}

To get insight into the dynamics of the CsPbI$_3$ perovskite structures at finite temperature, we first consider
the energy costs for structural transitions between these 24 structure models. To that end, we investigate the required energy barrier for changing from one spatial orientation of the $\gamma’$ structure to another
by a combination of changes of the tilt angles $\eta$ and $\theta$. 
There are only three different ways to describe the transitions between each two of the 24 variants. These three ways are illustrated in 
Fig.\ \ref{fig_stru_shift}(a). The respective intermediate transition states are shown in Fig.\ \ref{fig_stru_shift}(b) and the energy barriers for these transitions calculated using the CI-NEB method are plotted in Fig.\ \ref{fig_TS_energies}.

The first possibility is the inversion of the tilt angle $\eta$ (red arrows in Fig.\ \ref{fig_stru_shift}(a)). It requires a transition
path that reverts the tilt and thus passes through the $\beta’$ structure (with $\eta=0$) as
transition state,  c.f.\ left panel of Fig.\ \ref{fig_stru_shift}(b). Consequently, the calculated energy barrier for this process is the energy difference
of 0.41 eV between $\gamma’$ and $\beta’$ structures. As outlined above, the orientation of the tilt
angle $\eta$ is coupled to the direction of the off-center shift of Cs ions on A sites, i.e., the latter
is inverted together with the changed $\eta$.

The second possibility is the inversion of the tilt angle $\theta$ (blue arrows in Fig.\ \ref{fig_stru_shift}(a)). It requires a transition
through the $\beta'$ structure with $\theta=0$ while leaving $\eta$ unchanged, c.f.\ middle panel of Fig.\ \ref{fig_stru_shift}(b). The calculated energy barrier for this process is 0.29 eV.

Third, structural variants can be formed by inverting both $\theta=0$ and $\eta$ (black arrow in Fig.\ \ref{fig_stru_shift}(a)). This
transition passes through the symmetric $\alpha'$ structure (with $\theta=0$ and $\eta=0$), c.f.\ right panel of Fig.\ \ref{fig_stru_shift}(b), with a calculated energy barrier of 1 eV. 
Note that without redefining the rotation axis for $\theta$, these transitions can also be described as a combined
two-step procedure of changing $\theta$ and $\eta$ individually. However, this is not possible
when shifting the $\theta$ rotation axis, i.e., all three transition modes are necessary to connect
all 24 structural variants.

This analysis allows to interpret the structure-energy relationship as follows: Both $\beta$’ and $\alpha$’ structures are superpositions of $\gamma$’ structures and they mark the transition states. The octahedral tilts and off-center shifts of the Cs ions on the A sites are the determining factors (all the transition states shift the Cs into the center of the pseudo-cubic unit cell of the perovskite five-atoms formula unit), whereas the volume changes are small, with energetic stabilizations of -0.13~eV, -0.14~eV, and -0.03~eV per $2\times 2\times 2$ perovskite supercell for the $\alpha$, $\beta$, and $\gamma$ structures, respectively (see Fig.~\ref{fig_TS_energies}).

Note that our calculated energy barriers should be viewed as upper bounds for the actual energy costs for the structural transitions. These energies are obtained from periodic models, i.e., they correspond to a simultaneous change in the whole crystal. However, the more realistic dynamical changes will not be perfectly coupled. Although such realistic structural changes are likely to have some effect in the adjacent cells, the corresponding energy barrier should still not exceed the one calculated using periodic boundary conditions.

\subsection{Migration of Li atoms in the fast ion limit}

In the fast ion limit, we neglect the dynamical fluctuations of the CsPbI$_3$ perovskite structure. In this limit, Li atoms migrate fast through the host lattice without causing significant atomic displacements in the perovskite crystal. Considering this limit of fast and light atoms in lattices of slow and heavy atoms successfully provided a good physical understanding of states and barriers of interstitial hydrogen atoms in metals (cf., e.g., Refs.\ \cite{elsasser1991vibrational,kemali2000inelastic}).

\begin{figure}
\setlength{\unitlength}{1mm}
\begin{center}
\begin{picture}(85,115)(0,0)
\put(10,55) {\includegraphics[width=0.7\columnwidth]{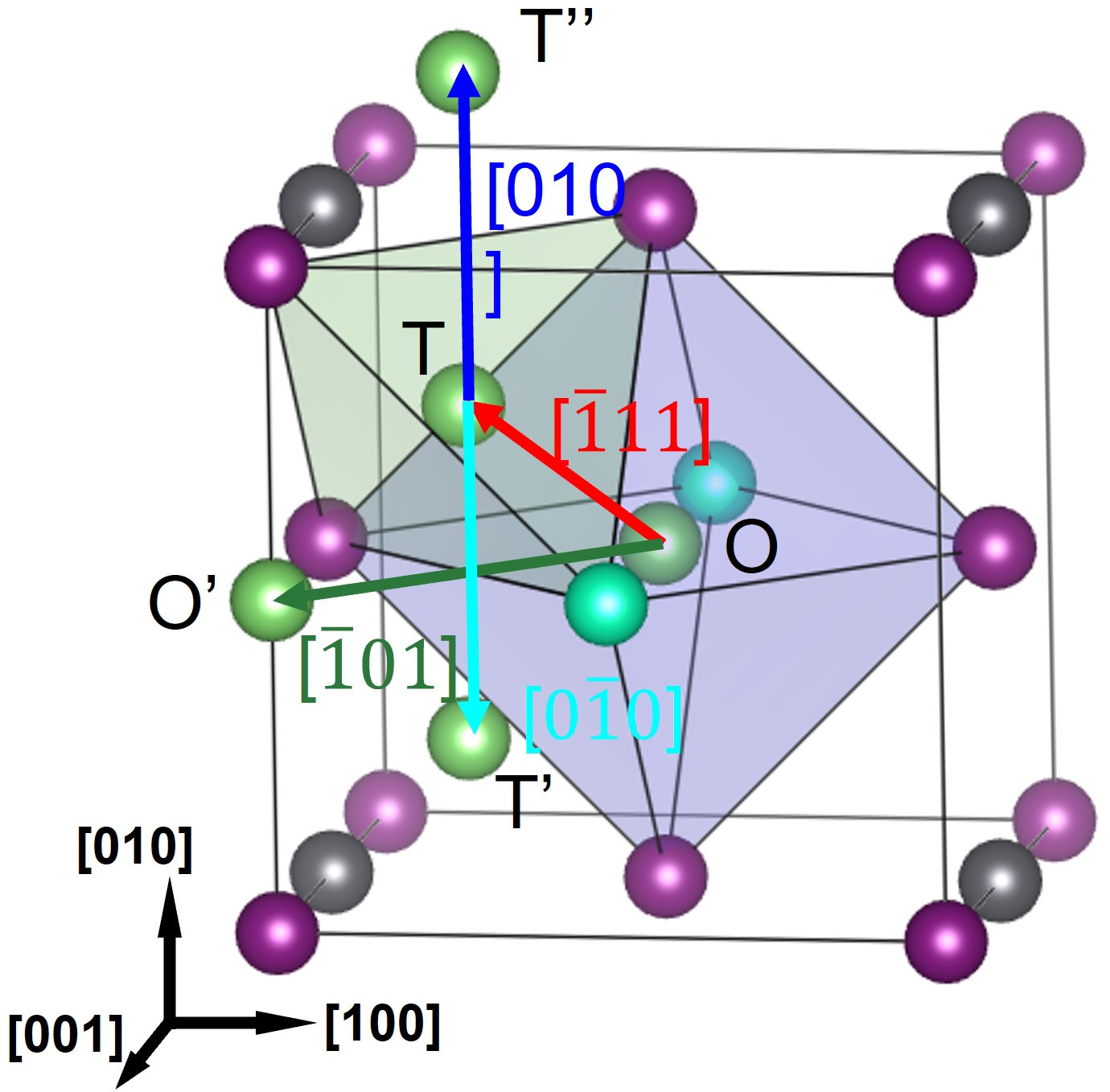}}
\put(0,0){\includegraphics[width=\columnwidth]{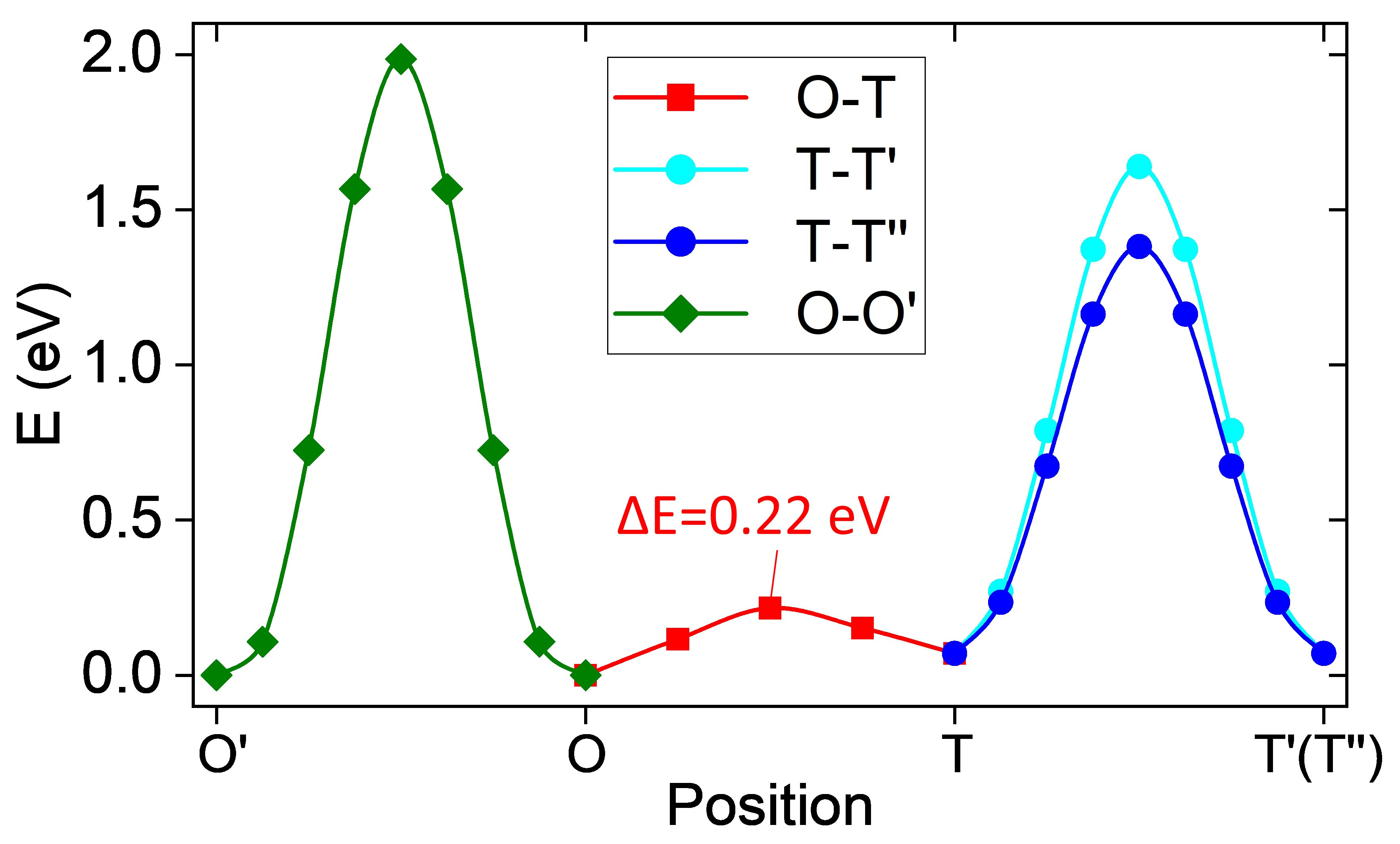}}
\put(0,52){(b)}
\put(0,110){(a)}
\end{picture}
\end{center}
	\caption{Migration paths of a Li ion in CsPbI$_3$ for the fast ion limit. Ions are indicated by cyan (Cs), dark grey (Pb), and purple (I) spheres. Interstitial Li positions are marked by green spheres. b) Energy barriers for the possible migration paths of a Li ion between adjacent interstitial sites. Energies are referenced with respect to Li occupying the most stable site (the optimized O site).}
	\label{fig_diffusion_alpha}
\end{figure}

Figure \ref{fig_diffusion_alpha} (a) displays the atomistic model of the cubic perovskite structure of the $\alpha$ phase with all the possible migration paths of interstitial Li atoms in the model of the fast ion limit. 
The insertion of a Li atom into the high-symmetry structure occurs on one of the three
octahedral or eight tetrahedral interstitial sites per unit cell \cite{polfus2016solubility,Yang2018lattice}. 
An octahedral site, denoted by O in the following, is located in the center of the blue
octahedron surrounded by four I atoms and two Cs atoms, a tetrahedral site, denoted by T, is in the center of
the green tetrahedron surrounded by three I atoms and one Cs atom. The migration paths of Li
from a T site to a next T' site and from a T site to another next T'' site are not the same because of the heterogeneous
crystal environment surrounding the interstitial Li atom. This heterogeneity leads to two
different migration paths TT' and TT'', while the sites T, T', and T'' themselves are symmetry equivalent.

The geometrically ideal O and T sites of Li ions in the cubic L1$_2$-type sublattice of Cs and I ions are slightly destabilized by minor spatial off-center shifts due to the
crystal surroundings. 
A small stabilization of -0.08 eV is obtained due to an off-center shift of the interstitial ion
along a [110] direction away from the geometrically ideal O-site location. Thus, each
of the three O sites becomes four-fold degenerate. Energy barriers for migration of Li
among the four-fold degenerate O sites (along a [100] or a [110] direction) are only 0.05~eV and 0.08~eV, respectively. The geometrically ideal T sites are also slightly destabilized by -0.04~eV due
to a small spatial off-center shift of the interstitial ion. In any case, O sites are slightly more stable than T sites
for a concentration of x=1/8 Li atoms per formula unit. Thus the stabilization of small off-center displacements does
not qualitatively change the interstitial O and T sites and the migration paths between them. The four-fold degenerate, off-centered O sites
can be treated as one ``effective'', centered O site due to the very small energy barriers
between them.

Figure \ref{fig_diffusion_alpha}(b) displays the energy profiles along all the possible Li migration paths between effective O and T sites. The energy profiles are obtained for the Li atom moving in straight lines from an initial interstitial site to a final interstitial site. The energy landscape for moving Li ions is characterized by just four migration paths. The migration path OO’ has the highest
energy barrier of approximately 2.0~eV, while the migration path OT has a barrier of only 0.22~eV. The transition-state energy of the MEP obtained from a CI-NEB calculation confirms the value of the OT
migration barrier. The migration paths TT' and TT'' have similar energy barriers
of approximately 1.5~eV. The OT migration paths provide a periodic three-dimensional network for long-range diffusion of interstitial Li ion in CsPbI$_3$ with a low energy barrier of 0.22~eV.

The PBE result for the OT migration barrier is checked and confirmed to be reasonable by recalculation using the SCAN meta-GGA functional. This functional improves the structure description of CsPbI$_3$ \cite{kaczkowski2021vibrational}. The lattice constant for the $\alpha$ structure is decreased to 6.31 Å with SCAN meta-GGA, which is 0.09 Å less than with PBE. The O site is 0.04 eV more stable than the T site, similar to the 0.07 eV difference obtained with PBE. The energy barrier from O to T is 0.24 eV, only 0.02 eV higher than the PBE results.

\subsection{Migration of Li atoms in the slow ion limit}

In the slow ion limit, the Li ions interact with the host-crystal atoms in a distorted $\gamma$' structure much more than in the high-symmetry $\alpha$ structure of the fast ion limit. In this slow ion limit, we allow displacements of the host-crystal atoms reacting to the presence of the migrating interstitial Li ion, i.e., we determine the static energy minima with respect to the positions of all the host-crystal surrounding an interstitial Li atom sitting on or moving between interstitial sites. 

\begin{figure}
	\centering
	\includegraphics[width=\columnwidth]{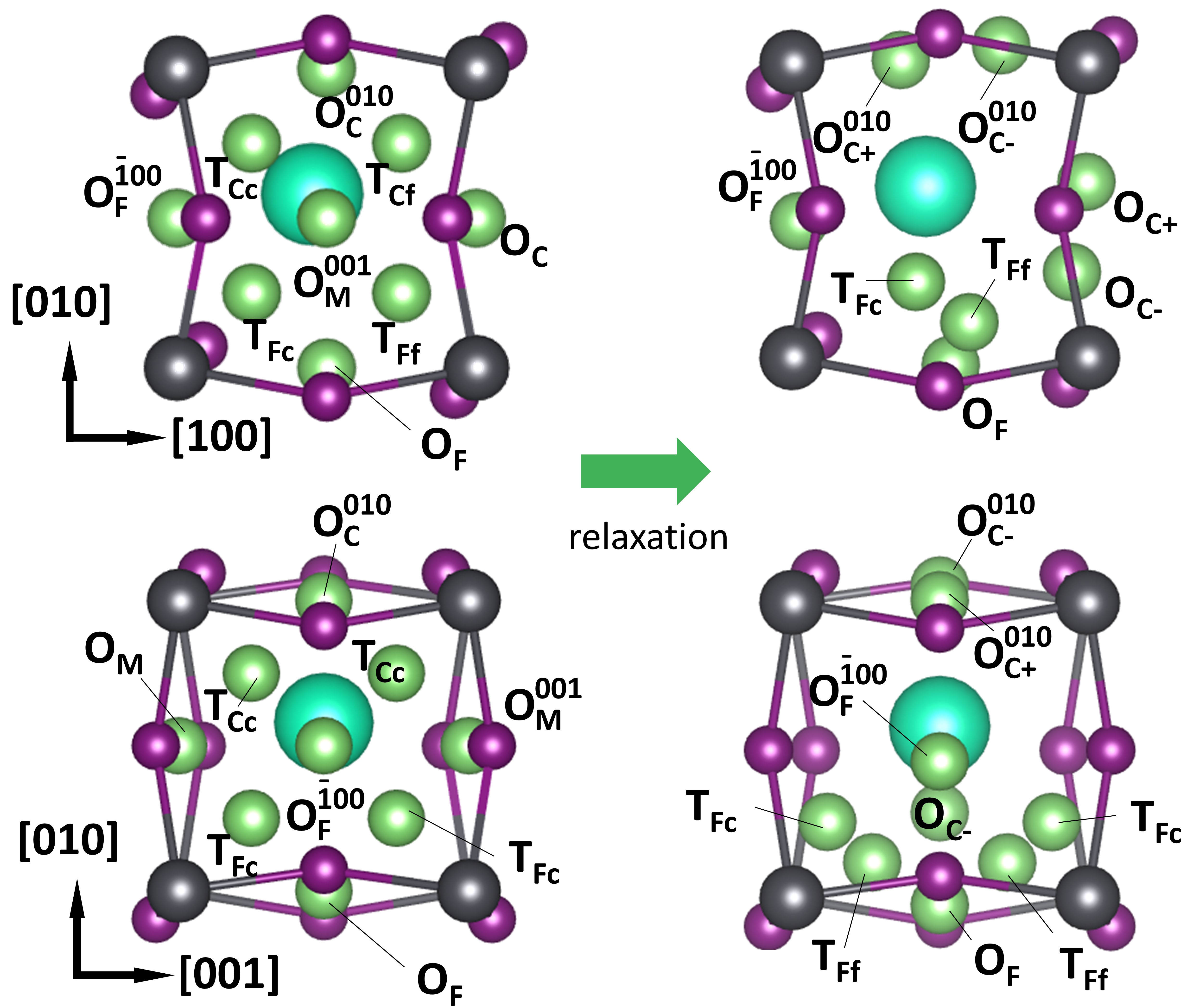}
	\caption{Li ions in the $\gamma$’ structure at their geometrical interstitial high-symmetry positions (left) and the structurally relaxed interstitial positions (right).
	In the $\gamma$’ structure we label the interstitial sites for Li ions by subscripts telling that they are closer to (C and c) or further from (F and f) the nearest Cs ion on the A site of the respective unit cell. For the T sites, the pairs of uppercase and lowercase letters distinguish their distances to the A site in the [010] and [100] directions, respectively.
	Furthermore, we indicate the locations of the interstitial sites with respect to a pseudo-cubic reference unit cell with a respective superscript label. For example, the label 010 indicates an interstitial site inside the next neighboring unit cell along [010]. The label of the reference cell, 000, is omitted for clarity.
	}
	\label{sites_gamma}	
\end{figure}


\begin{table}
	\centering
	\begin{tabular}{lccccc}
	\hline \hline
	&$\;$ T$_{\rm Ff}$ $\;$
	&$\;$ T$_{\rm Fc}$ $\;$
	&$\;$ O$_{\rm C-}$ $\;$
	&$\;$ O$_{\rm C+}$ $\;$
	&$\;$ O$_{\rm F}$  $\;$\\ 
	\hline
	$d_{\rm Cs-Li}$ [\AA]            & 3.97 & 3.83 & 4.01 & 3.79 & 3.73 \\
	$\Delta r_{\rm Cs}$ [\AA]        & 0.67 & 0.66 & 0.65 & 0.66 & 0.70 \\
	$\Delta_{\rm Pb-I-Pb}$ [$^\circ$]&-28.7 &-28.5 &-27.7 &-27.2 &-26.3 \\
	$E_{\rm abs}$ [eV]             &-0.26 &-0.22 &-0.16 &-0.05 &-0.02 \\
	\hline \hline
	\end{tabular}
	\caption{Stable interstitial positions of Li in Li$_x$CsPbI$_3$ with $x$=1/8: the absorption energy $E_{\rm abs}$,
	 the distance of Li to its nearest Cs neighbor $d_{\rm Cs-Li}$, the Cs off-center displacement 
	 $\Delta r_{\rm Cs}$, and the change in the  Pb-I-Pb bond angle $\Delta_{\rm Pb-I-Pb}$ are given.}
	\label{tab_stru}
\end{table}

\begin{figure}
	\centering
	\includegraphics[width=\columnwidth]{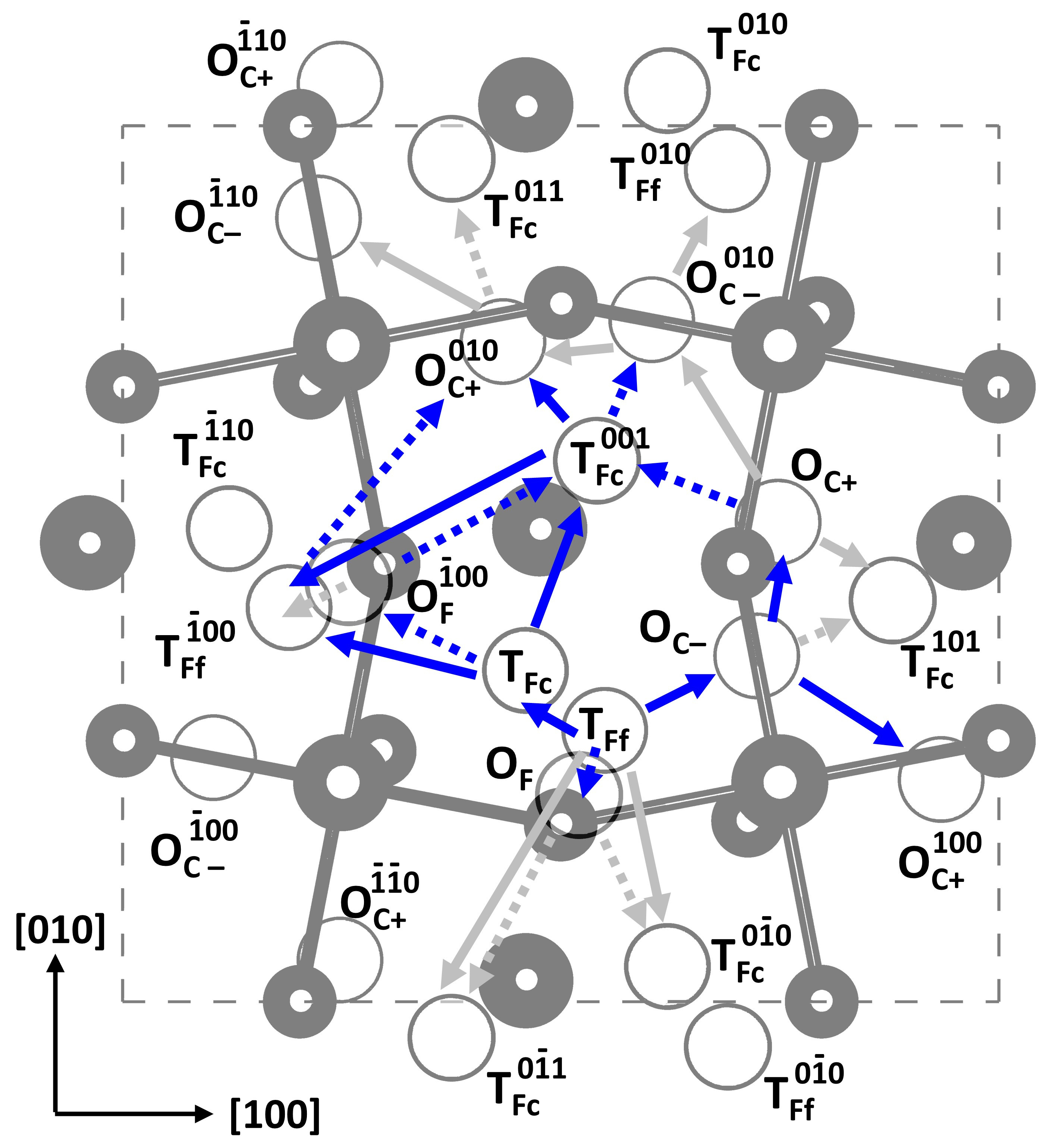}	
	\caption{
Atomic jumps (indicated by blue arrows) connecting adjacent interstitial Li sites (shown as open black circles) in the $\gamma$' structure. Dashed arrows indicate atomic jumps with higher energy barriers than other possible paths in their vicinity.	
Gray arrows indicate symmetry equivalent jumps to those shown in blue. 
(For the meaning of subscript and superscript labels see the caption of Fig. \ref{sites_gamma}.)}
	\label{migration_gamma}
\end{figure}

Figure \ref{sites_gamma} displays the stable interstitial sites for Li ions in the $\gamma$' structure of CsPbI$_3$ at a Li concentration $x$=1/8. Due to the lower symmetry of the $\gamma$’ structure, the three O and eight T sites from the $\alpha$ structure  are no longer symmetry-equivalent. We denote the interstitial sites with respect to being close to (C and c) or far from (F and f) the respective Cs atom on the A site of the respective unit cell. 
The eight T sites in each unit then split into four pairs of sites T$_{\rm Ff}$, T$_{\rm Fc}$, T$_{\rm Cf}$ and T$_{\rm Cc}$. The three O sites are distinguished as the non-equivalent O$_{\rm F}$ and O$_{\rm C}$, with respect to being far from or close to the Cs ion, and the third one O$_{\rm M}$ with middle distance (neither far nor close) to the Cs ion. After relaxation of these seven distinct interstitial sites, shown in the left two graphics of Fig. \ref{sites_gamma}, only five sites, shown in the right two graphics of Fig. \ref{sites_gamma}, are found to be stable, namely T$_{\rm Ff}$, T$_{\rm Fc}$, O$_{\rm C-}$, O$_{\rm C+}$, and O$_{\rm F}$. The “Close” sites T$_{\rm Cc}$ and T$_{\rm Cf}$ are unstable and structural relaxation shifts the Li ion to one of the stable interstitial positions.

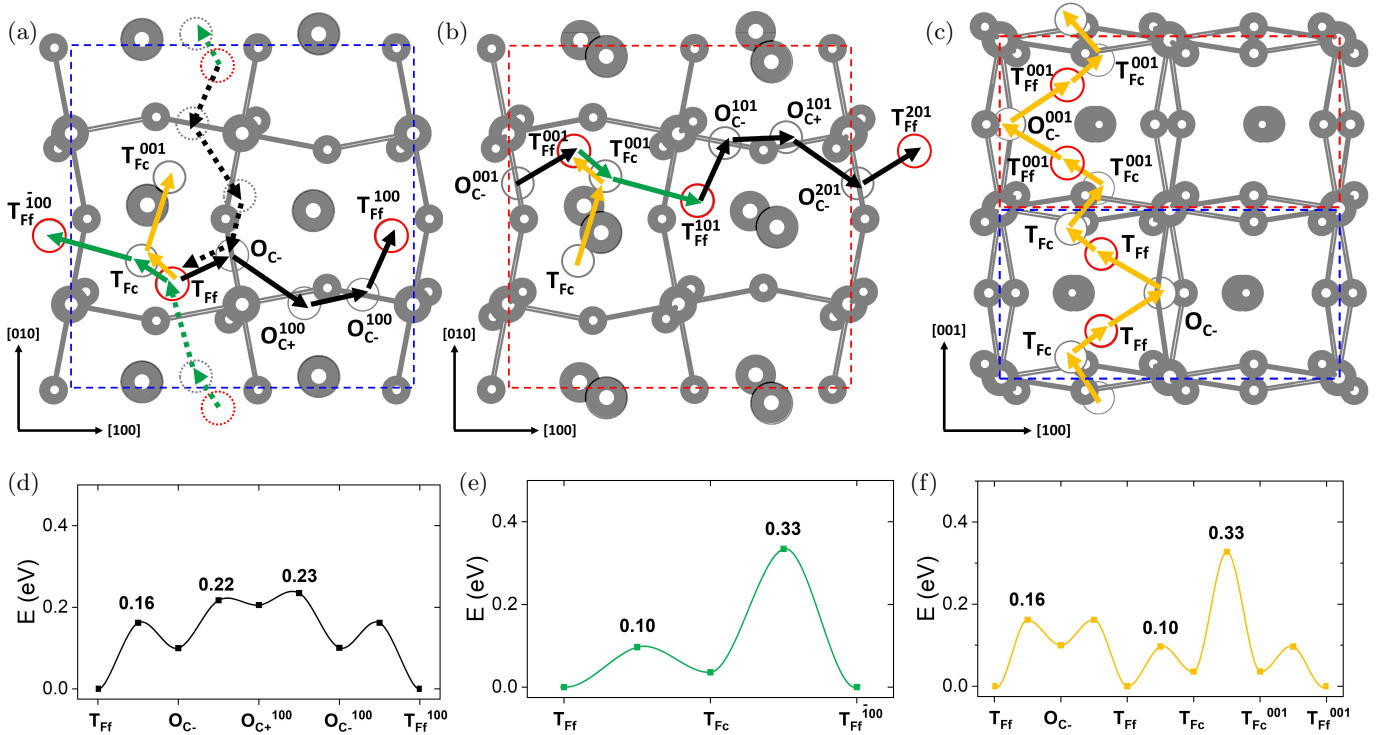
\begin{figure*}
\setlength{\unitlength}{1mm}
\begin{center}
\begin{picture}(180,95)(0,0)
\put(  0,0){\includegraphics[width=0.68\columnwidth]{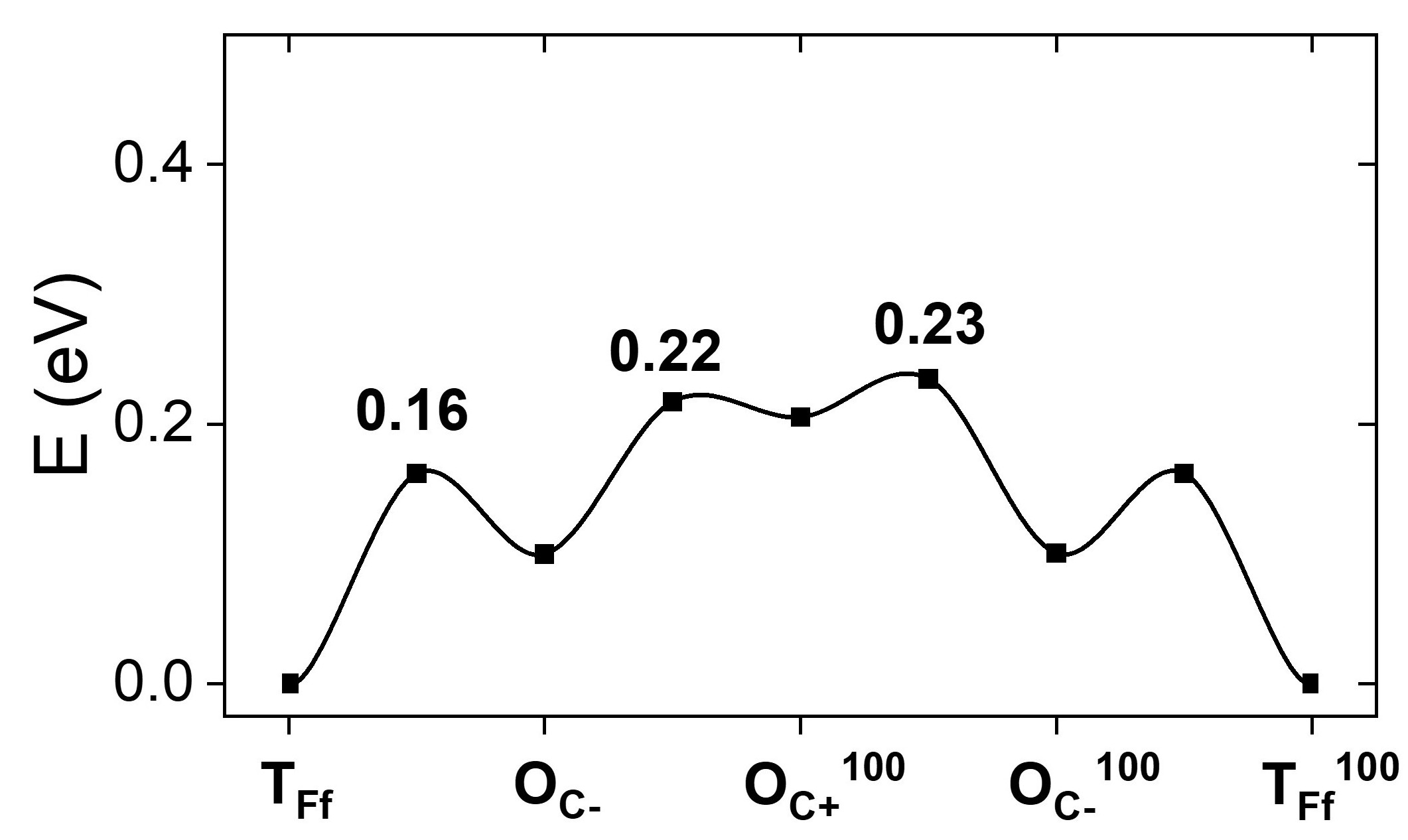}}
\put( 60,0){\includegraphics[width=0.68\columnwidth]{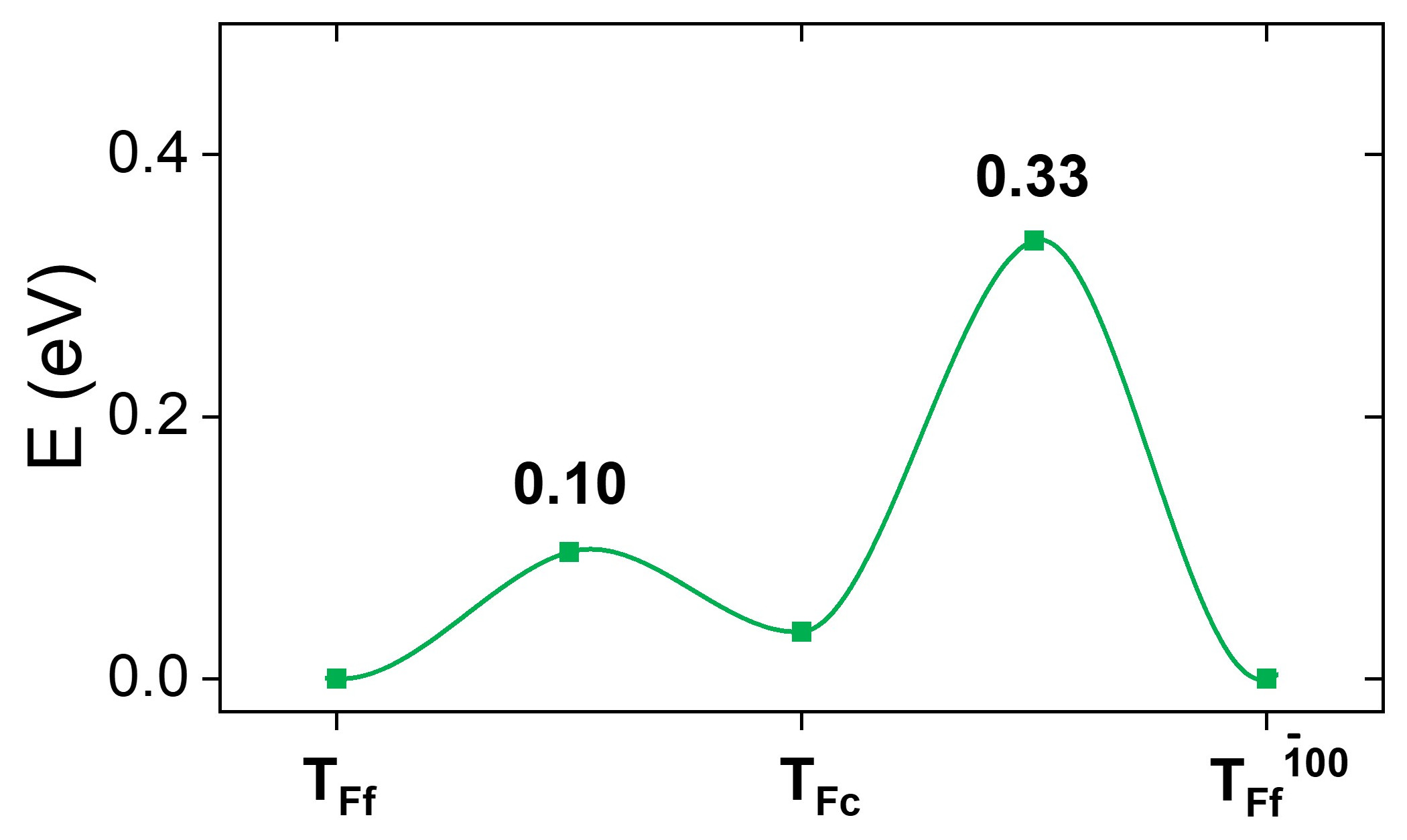}}
\put(120,0){\includegraphics[width=0.68\columnwidth]{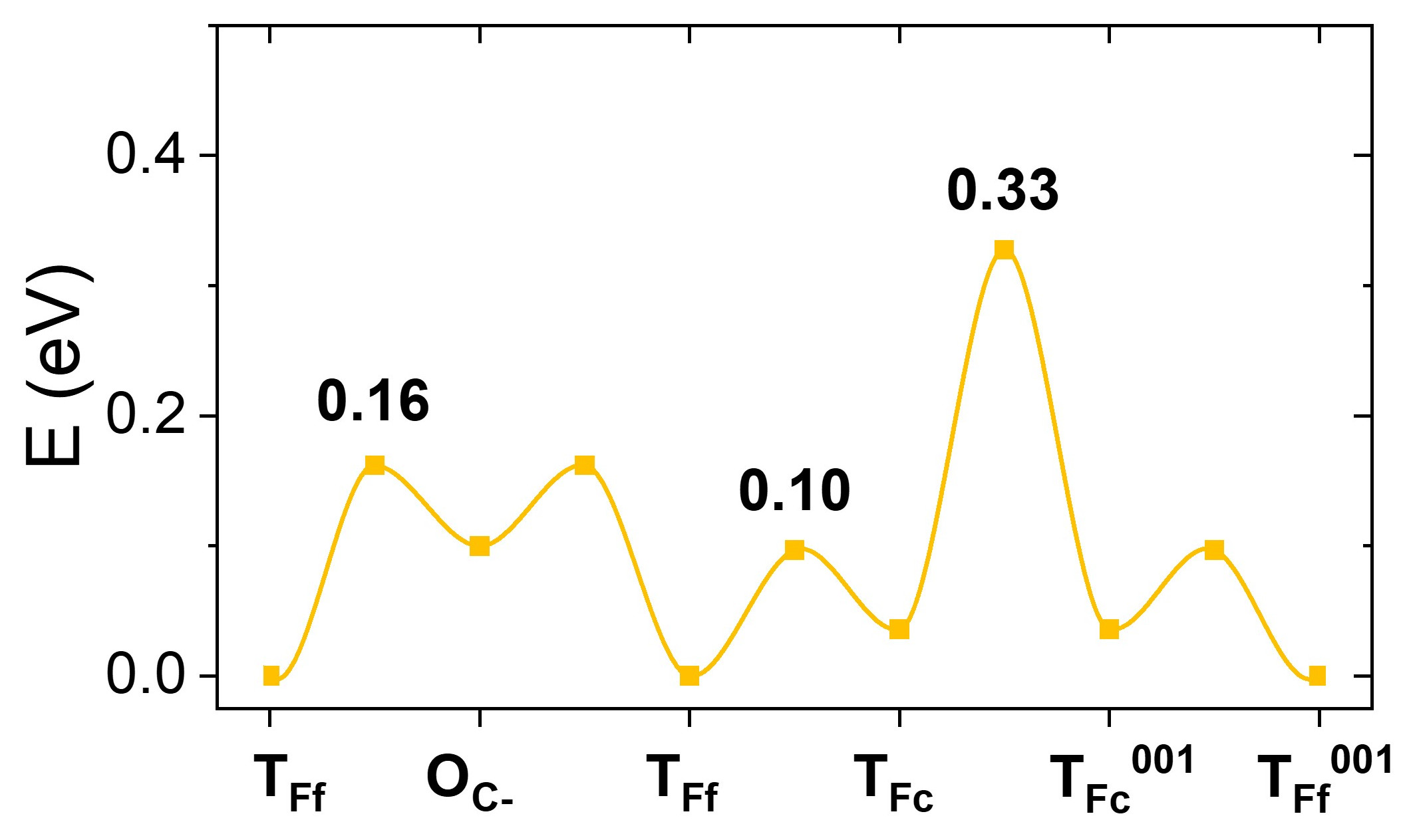}}
\put(  0,33){(d)}
\put( 60,33){(e)}
\put(120,33){(f)}
\put(  0,40){\includegraphics[width=0.67\columnwidth]{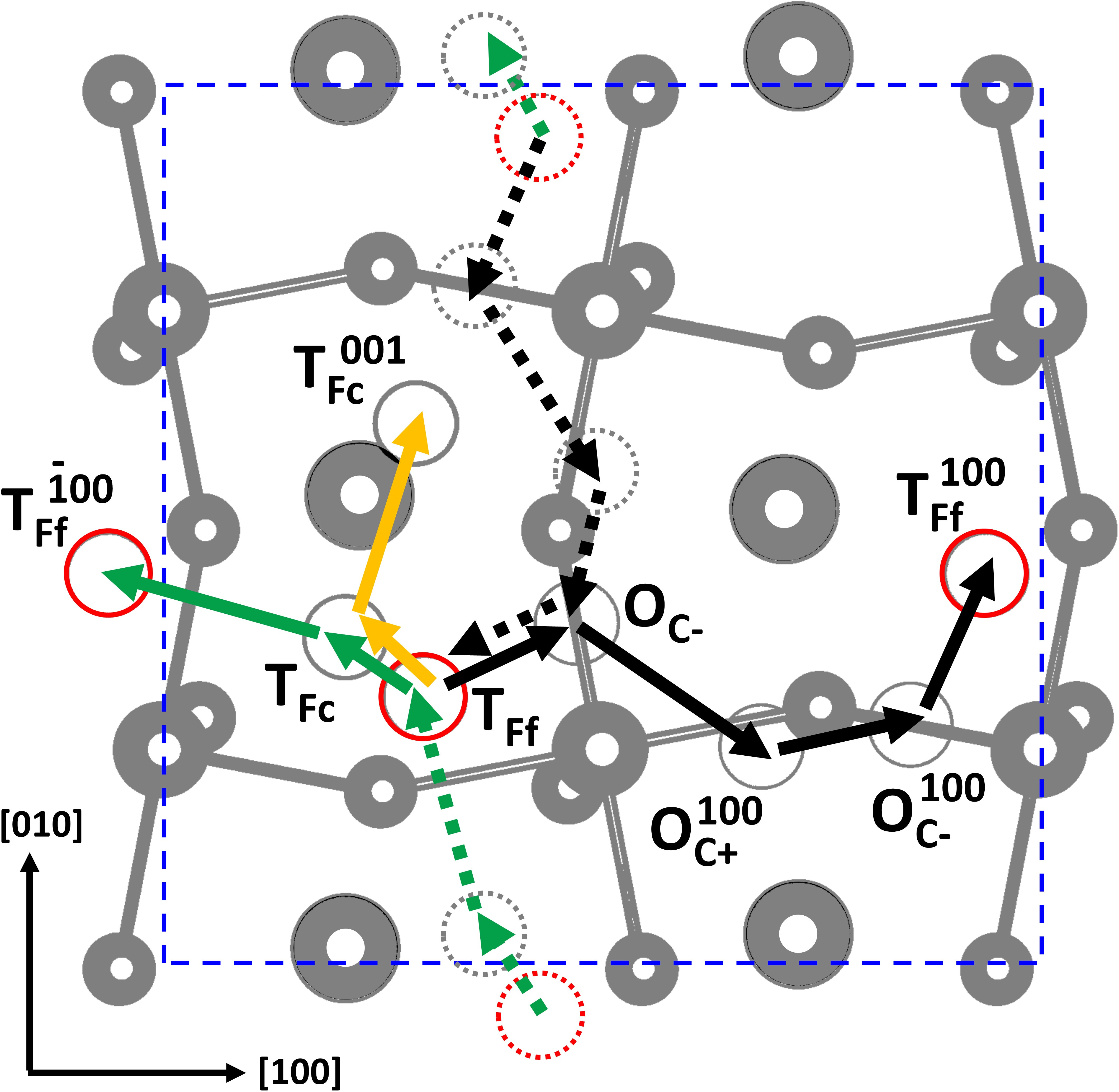}}
\put( 58,40){\includegraphics[width=0.75\columnwidth]{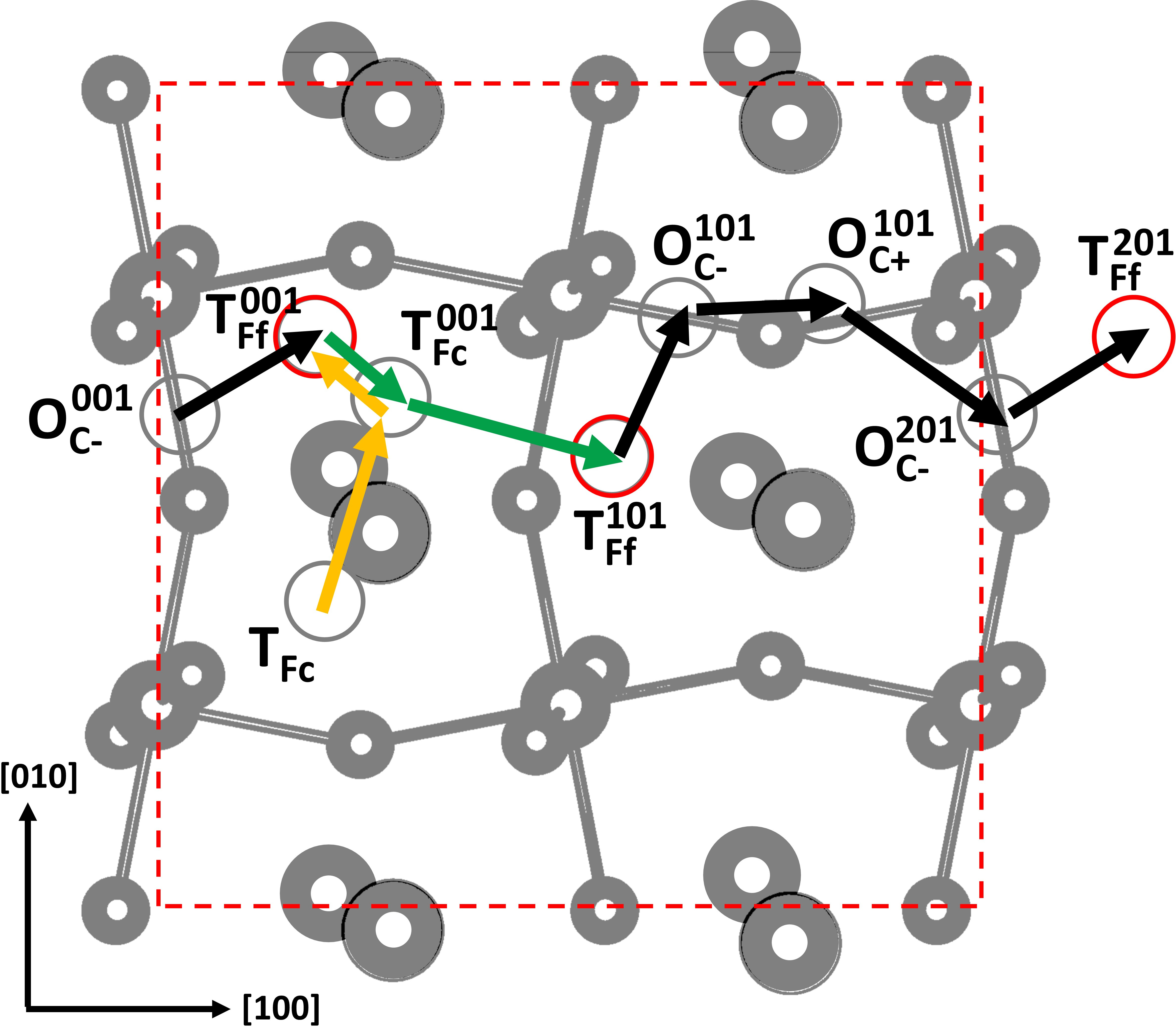}}
\put(123,40){\includegraphics[width=0.67\columnwidth]{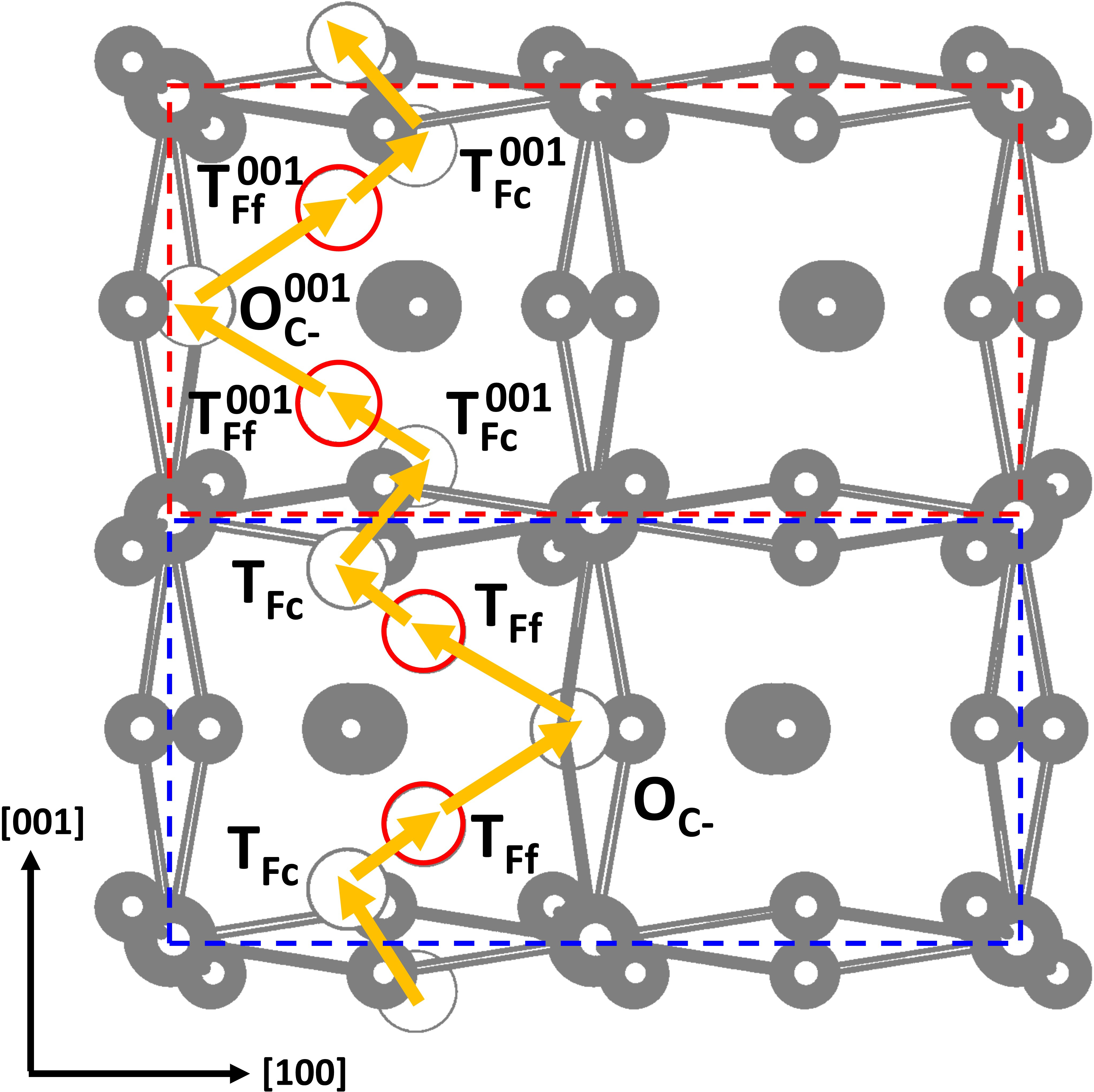}}
\put(  0,93){(a)}
\put( 57,93){(b)}
\put(122,93){(c)}
\end{picture}
\end{center}
	\caption{(a)-(c) Migration paths from one T$_{\rm Ff}$ site to its symmetry equivalent next neighbor T$_{\rm Ff}$ for the slow ion limit, displayed from three different perspectives. The T$_{\rm Ff}$ sites are marked by red circles, all other interstitial sites by gray circles. Black, green, and yellow arrows mark the three migration paths with the lowest energy barriers along [100],[-100], and [001] directions. The dashed arrows represent symmetry-equivalent paths.  The colored dashed frames indicate the different layers of the perovskite structure.(d)-(f) The energy profiles for the black, green, and yellow migration paths.
	(For the meaning of subscript and superscript labels see the caption of Fig. \ref{sites_gamma}.)}
	\label{fig_diffusion_gamma}
\end{figure*}

Table~\ref{tab_stru} lists the absorption energy ($E_{\rm abs}$) of Li in CsPbI$_3$, the distance of Li to its nearest Cs neighbor ($d_{\rm Cs-Li}$), the Cs off-center displacement ($\Delta r_{\rm Cs}$), and the change of the Pb-I-Pb bond angles ($\Delta_{\rm Pb-I-Pb}$) of all five energetically stable interstitial sites. 
Among these interstitial sites, T$_{\rm Ff}$ is the most stable one with an absorption energy of -0.26~eV and the least stable one is O$_{\rm F}$ with only -0.02~eV.

Here we observe a different hierarchy of stability between O and T sites, as compared to the high-symmetry $\alpha$ structure, where the O site is slightly more stable than the T site. The obtained absorption energies of interstitial Li in the $\gamma$’ phase can be rationalized as follows: First, due to the varying interactions of the Li ion and the surrounding host-crystal ions, in particular the A-site Cs ions, the interstitial positions (and the crystal surroundings) are subject to structural rearrangements. Proximity of Li and Cs leads to unstable structures and an additional shift of Cs due to cation-cation repulsion. Consequently, the energetic ordering of the not symmetry-related T and O sites coincides with the distance of Li and the nearest Cs ion ($d_{\rm Cs-Li}$), with less stable sites found for shorter $d_{\rm Cs-Li}$. Then comparing O and T sites, the latter are the more stable interstitial sites for Li ions in the $\gamma$’ structure. This can be explained with shifts of Li ions away from T-site centers, as compared to the analogous off-center shifts in the $\alpha$ structure. As a result of the shifts, away from Cs sites and towards Pb sites, Li atoms on T sites in the $\gamma$’ structure are surrounded with four I atoms (compared to three I atoms in the $\alpha$ structure). Lastly, Pb-I-Pb bond angles are also changed by the presence of the Li cation, due to its attraction to surrounding I anions. $\Delta_{\rm Pb-I-Pb}$ is the average change of Pb-I-Pb bond angles, as compared to the $\alpha$ structure. A larger tilt of the Pb-I-Pb bond angles coincides with a lower absorption energy. However, this is a byproduct of stronger Li-I bonds, since the most stable $\Delta_{\rm Pb-I-Pb}$ is -27.1$^\circ$ in the $\gamma$’ structure without Li.

In the following, we investigate the possible migration paths between interstitial sites within the unit cell and between adjacent unit cells in order to construct and analyze long-range diffusion pathways for interstitial Li atoms in the slow ion limit. Figure \ref{migration_gamma} displays all considered atomic jumps between adjacent interstitial sites.
The dashed arrows mark alternative paths with higher energy barriers.

To identify possible pathways for three-dimensional long-range diffusion of Li atoms, we start from the most stable interstitial Li position, T$_{\rm Ff}$. From here we construct continuous paths that connect stable interstitial sites. When we exclude the least stable O$_{\rm F}$ site, the long-range Li-diffusion pathways are given by sequences of atomic jumps between neighboring T$_{\rm Ff}$, T$_{\rm Fc}$, O$_{\rm C+}$ and O$_{\rm C-}$ sites. 
Note that we are here discussing migration of Li atoms through one static orientation of the $\gamma$’ structure. Due to symmetry, the migration paths in the [100] and [010] directions are equivalent. Thus we investigate the long-range diffusion pathways in [100] and [001] directions. 

Figure~\ref{fig_diffusion_gamma} displays in panels (a)--(c) the three migration paths with the lowest energy barriers that connect the most stable interstitial positions, T$_{\rm Ff}$, of neighboring unit cells in all three crystallographic directions. The corresponding migration energy profiles are displayed in Figs.~\ref{fig_diffusion_gamma}(d)-(f). All other possible paths (along migration steps marked by dashed arrows in Fig.~\ref{migration_gamma}) have higher energy barriers and are, therefore, omitted in the following. The Li migration between neighboring T$_{\rm Ff}$ positions depends on the direction. The black and green arrows illustrate the migrations along [100] or [-100], i.e.\ from T${\rm _{Ff}}$ to T${\rm _{Ff}}^{100}$ and T${\rm _{Ff}}^{-100}$, respectively. The critical migration step along the black path in [100] direction in Fig.~\ref{fig_diffusion_gamma} occurs from O${\rm _{C+}^{100}}$ to O${\rm _{C-}^{100}}$, with an energy barrier of 0.23~eV. The critical migration step along the green path in [-100] direction in Fig.~\ref{fig_diffusion_gamma} involves a jump from T${\rm _{Fc}}$ to T${\rm _{Ff}}^{-100}$ with a higher energy barrier of 0.33~eV. The dashed black and green arrows represent the symmetry equivalent migration paths in the [010] and [0-10] directions. Furthermore, the long-range diffusion along [001] is primarily determined by the migration step from T${\rm _{Fc}}$ to T${\rm _{Fc}}^{001}$ with an energy barrier of~0.33 eV. The blue and red dashed frames emphasize the distinct layers of perovskite formula-unit cells, which contribute to the construction of a comprehensive three-dimensional network of pathways for long-range Li diffusion within the entire simulation box, based on the chosen configuration of the $\gamma$' structure. 

Along the three pathways displayed in Figure~\ref{fig_diffusion_gamma}, macroscopic long-range diffusion of Li ions can be achieved with an energy barrier of only 0.23 eV by following exclusively segments of the black path. (In Fig. \ref{fig_diffusion_gamma} this is a a zig-zag path, marked by the solid and dashed black arrows, along the [1-10] direction). Along all other pathways that include segments of the green or yellow paths, diffusing Li ions have to pass a higher energy barrier of 0.33 eV.

\section{Discussion} \label{discussion}


In this study, we propose two limiting models to capture the different aspects of Li diffusion in the dynamical crystal structure of the $\alpha$ phase of CsPbI$_3$. The fast ion limit model represents the configuration at the highest saddle point of the vibrating CsPbI$_3$ crystal, while the slow ion limit model captures the stable configurations at the local energy minima.  
The study of these two limiting models allows us to derive upper and lower bounds for the effect of dynamic vibrations, whereas the computational resources required to implement the true dynamic processes in DFT calculations for this investigation would be very demanding \cite{gebhardt2021efficient}.

In the model of the fast ion limit, the $\alpha$ structure, long-range diffusion in all directions can be constructed from O-T migration steps with a low energy barrier of 0.22~eV. 
The migration of interstitial Li in the model of the slow ion limit, the $\gamma$' structure, is more complicated. The lowest energy barrier of 0.23~eV for diffusion is practically the same as the barrier of 0.22~eV of the fast ion limit. But along most directions there is a higher energy barrier of 0.33~eV, which has to be overcome in the slow ion limit. 
Depending on the migration direction, the potential energy barriers for long-range diffusion can be either 0.23~eV or 0.33~eV, cf. Fig.~\ref{fig_diffusion_gamma}.

Note that the observed direction dependence of diffusion is only valid for a specific variant of the $\gamma$’ structure. Due to the dynamical transitions between all 24 variants of the crystal, the path with the lowest energy barrier does not have a specific direction, but can point in any of the spatial directions. 


So far we only deal with one of the 24 variants of the $\gamma$’ structure in the slow ion limit. However, the interstitial sites and diffusion pathways for Li atoms in all variants are symmetry equivalent. But the cooperative transitions between the distorted structure variants upon Li insertion have not yet been addressed. In the discussion of the dynamic nature of the perovskite structure at finite temperature, we determined an activation energy of at least 0.29 eV for the required reverting of tilt angles of the $\gamma$’ structure without interstitial Li ions (similar to 0.23~eV for the retilting of the $\gamma$ phase calculated by Klarbring \cite{klarbring2019low}). Since this activation energy for the MEP of the transformation (marked by the blue arrows in Fig.~\ref{fig_stru_shift}) is in the energy range required for the migration of Li ions, we further investigated whether the presence of Li ions favors this transformation process. Hence, the question is, whether the expected dynamical structure changes from one $\gamma$’ structure variant to another can lead to migration paths for Li ions with lower energy barriers. 
The transition via the $\beta_\eta$’ structure, by an inversion of the angle $\theta$ and the shift of A-site Cs ions, has the smallest energy barrier. Therefore, we investigated the effect of interstitial Li at the T$_{\rm Fc}$ position on this transition. The resulting energy barrier is 0.45~eV, which is 0.16~eV higher than the energy barrier without Li (see Fig.\  \ref{fig_TS_energies}). The other two possible transition paths marked by the red and black arrows in Figure \ref{fig_stru_shift} have higher energy barriers and are therefore not further considered. This suggests that the Li diffusion and the structural reorientation are more likely to occur separately. 

Hence, we infer: i) the presence of Li ions does not enhance the kinetics of the dynamical structural changes of the host perovskite structure of CsPbI$_3$, and ii) the barriers for the migrating Li ions obtained in one specific $\gamma$’ variant, carried out as described above, are not expected to be lowered by taking the dynamical structural changes of the host lattice into account. Therefore, our study based on static crystal structures should provide a reliable upper limit for the lowest diffusion barriers of Li through CsPbI$_3$ even at finite temperatures.

The two extreme cases of Li migration in CsPbI$_3$ are represented by the fast ion limit, where there is no adaption of the crystal to the Li ions, and the slow ion limit, where there is complete relaxation of the crystal around the Li ions. Therefore, the true energy barrier for Li migration should lie between the energy barriers obtained for these two cases, namely 0.22~eV and 0.33~eV.

The recent DFT studies on the transition between symmetrical and distorted cubic structures are consistent with our findings. Yang et al. \cite{yang2017spontaneous} suggested that the transition from the high symmetry $\alpha$ structure to distorted structures occurs without encountering energy barriers, which is in good agreement with our calculations (cf. Fig.~\ref{fig_stru_shift}). They estimated the energy difference between $\alpha$ and distorted structures to be at least 0.51 eV (for crystal cells containing eight formula units) based on harmonic phonon modes. Marronnier et al. \cite{marronnier2017structural} obtained similar results through calculations of soft phonon modes, providing a more comprehensive energy landscape of dynamic vibrations. However, the actual energy differences are expected to be higher than those calculated using the phonon model. Klarbring \cite{klarbring2019low} investigated the energy difference between the two extreme structures using the CI-NEB method and obtained an energy difference of 1.10~eV, which agrees well with our value of 1.01~eV. Additionally, the reported barrier of at least 0.23~eV for retilting the Pb-I octahedra in the disordered cubic phase is similar to the activation energy of at least 0.29~eV required for retilting the $\gamma$' structure in our calculations. Thus, we are confident that the models of the two limits are representative for the realistic $\alpha$-CsPbI$_3$ and useful for studying the location and migration of interstitial Li ions in CsPbI$_3$.

\begin{table*}
	\begin{ruledtabular}
	\begin{tabular}{ccccccc}
    &Li$_{1/8}$CsPbI$_3$
	&LiCoO$_2$\cite{koyama2012defect,ning2014strain}
	&LiFePO$_4$\cite{shi2016density,hoang2011tailoring}
	&Li$_4$Ti$_5$O$_{12}$\cite{duan2015tailoring,ziebarth2014lithium}
	&Li$_{1/72}$TiO$_2$\cite{yeh2018first}
	&Li$_{1/8}$Nb$_2$O$_5$\cite{jing2021engineering}\\ 
	\hline
	E$\rm _{abs}$     [eV] &-0.26&-0.54&-0.42&-0.62      &-1.26 to -1.41&-3.6\\			
	E$\rm _{barrier}$ [eV] & 0.22 to 0.33& 0.39& 0.48& 0.30 to 0.48& 0.30 to 0.51 & 0.74 \\			
		\end{tabular}
	\end{ruledtabular}
	\caption{The absorption energy E$\rm _{abs}$ and migration barrier E$\rm _{barrier}$ of Li ions in Li$_{1/8}$CsPbI$_3$ calculated in the present work, in comparison to results for a selection of commercially successful Li-ion battery materials from comparable DFT studies. 
	}
	\label{commercial_comparison}
\end{table*}

Finally, we estimate the applicability of Li$_{x}$CsPbI$_3$ as an electrode material for Li-ion photo-batteries. 
Therefore, we compare the formation energies and migration barriers for interstitial Li atoms in our halide perovskite system with data for some presently commercialized or well-studied lithium storage materials. Three decades ago, Sony Co., Japan, developed the first commercial Li-ion battery and used the layer-type compound lithium cobalt oxide (Li$_x$CoO$_2$) for the cathode \cite{nagaura1990progress}. Since then, prosperous efforts were made to optimize this electrode material in terms of low cost, improvement of energy and power density, and non-toxicity \cite{reddy2013metal,mi2005situ,park2010based}.
For now, the lithium storage in electrodes of Li-ion batteries are based on several mechanisms, i.e., a Li intercalation-deintercalation reaction, alloying-dealloying reaction, or conversion (redox) reaction \cite{reddy2013metal}.

Since our system is based on intercalation and deintercalation of interstitial Li ions, we only compare to electrode materials with a similar Li uptake mechanism. 
LiFePO$_4$ consists of adequate raw materials, is also non-toxic, and has good cyclability. Li$_x$Ti$_5$O$_{12}$ (from $x$=4 to $x$=7) is also a promising and well-studied zero-strain material, due to the almost unchanged volume during the Li intercalation-deintercalation process \cite{ziebarth2014lithium}. In addition, the versatile TiO$_2$ has been examined in the literature for its good cyclability. In general, Li$_x$TiO$_2$ in nano-structured forms can deliver Li stably and near its theoretical capacity limit for long time \cite{yang2009nanostructures,froschl2012high}. Besides these three materials,  Li$_x$Nb$_2$O$_5$ should also be noticed because of its high Li capacity \cite{kodama2006electrochemical,kumagai1999thermodynamics,lantelme2000study}. Table~\ref{commercial_comparison} gives the formation energies and Li diffusion barriers of the present Li$_x$CsPbI$_3$ with $x = 1/8$, in comparison to the above mentioned, established electrode materials. All of the data in Table~\ref{commercial_comparison} are calculated using DFT methods. The migration barrier of Li in CsPbI$_3$ is similar to those of the other materials, even a bit lower, which may indicate a very good diffusion behavior. The absorption energy for Li in the present halide perovskite system is negative, but less negative than those of other listed materials. This means that the Li atoms can be put into the CsPbI$_3$ crystal. However, the combination of Li with this crystal is less stable than with the other materials. Altogether, Li$_x$CsPbI$_3$ is theoretically a suitable system for an electrode of a Li-ion battery. 

\section{Summary} \label{summary}

In summary, we have studied the location and migration of interstitial Li ions in the halide perovskite CsPbI$_3$ as a representative model compound for a halide-perovskite photo-battery material. We have considered two scenarios: the fast-ion limit with Li ions moving in the rigid $\alpha$ structure of CsPbI$_3$, and the slow-ion limit, where Li fully interacts with the flexible host crystal structure, leading to a distorted $\gamma$' structure that is closely related to the $\gamma$ phase of CsPbI$_3$. 
The higher symmetry structures are obtained as transition states when the octahedral tilts of the $\gamma$ phase are changed dynamically at finite temperatures, and we gave an estimate for the required activation energy barriers for these transitions.
Migration of Li ions in the fast-ion limit requires to overcome an energy barrier of 0.22~eV, with interstitial O sites being the most favorable interstitial sites for Li ions. In contrast to the fast-ion limit, in the slow-ion limit investigated using the $\gamma$’ structure, uptake of interstitial Li ion is significantly favored, leading to energetically stable structures for the investigated composition of Li$_{1/8}$CsPbI$_3$, in line with experimentally studied Li-containing halide perovskite compounds. 
The structural changes in the host crystal lead to a stabilization of T sites over O sites, too. Due to the lowered symmetry, initially equivalent interstitial Li-ion positions become finally non-equivalent and the network of migration paths becomes locally anisotropic. Despite the local complexities of Li$_{1/8}$CsPbI$_3$ in the $\gamma$’ structure, macroscopic long-range diffusion can be related to a minimal energy barrier for microscopic migration of Li ions, which is almost the same in the slow-ion limit as in the fast-ion limit. From our results and analyses we conclude that CsPbI$_3$ can be a suitable electrode material for a Li-ion photo-battery. Given the success of lead-based halide preovskites as photo-voltaic light harvesting materials, such compounds may also be suitable for integrated photo-battery devices to harvest, store, and deliver electrical energy.

\begin{acknowledgments}
This work was supported by the Deutsche Forschungsgemeinschaft (DFG, German Research Foundation) under Germany’s Excellence Strategy-EXC-2193/1-390951807 (LivMatS). We thank the State of Baden-Württemberg (Germany) through bwHPC for computational resources.
\end{acknowledgments}


%

\end{document}